\pdfoutput=1

\documentclass[preprint,3p]{elsarticle}

\usepackage{graphicx,amssymb}
\usepackage{epsf,psfig}

\journal{Ioannina, 5--8 September 2011}

\begin{document}

\begin{frontmatter}

\title{Order and chaos in a triaxial galaxy model with a dark halo component}

\author{Nicolaos D. Caranicolas}
\author{Euaggelos E. Zotos\corref{}}

\address{Department of Physics, \\
Section of Astrophysics, Astronomy and Mechanics, \\
Aristotle University of Thessaloniki \\
GR-541 24, Thessaloniki, Greece}

\cortext[]{Corresponding author: \\
\textit{E-mail address}: evzotos@astro.auth.gr (Euaggelos E. Zotos)}

\begin{abstract}

We study the regular or chaotic nature of orbits in a 3D potential describing a triaxial galaxy surrounded by a spherical dark halo component. Our numerical calculations show, that the percentage of chaotic orbits decreases exponentially, as the mass of the dark halo increases. A linear increase of the percentage of the chaotic orbits was observed, as the scale length of the dark halo component increases. In order to distinguish between regular and chaotic character of orbits, we use the total angular momentum $L_{tot}$, as a new indicator. Comparison of this new dynamical parameter, with other, previously used chaos indicators, shows that the $L_{tot}$ gives very fast and reliable results in order to detect the character of orbits in galactic potentials.

\end{abstract}


\end{frontmatter}

\section{Introduction}

We shall study the motion in a 3D composite galactic model described by the potential
\begin{equation}
{{V}_{t}}\left( x,y,z \right) = {{V}_{g}}\left(x,y,z \right) + {{V}_{h}}\left(x,y,z \right),
\end{equation}
where
\begin{equation}
{{V}_{g}}\left(x,y,z \right) = \frac{\upsilon _{0}^{2}}{2}\ln \left[ {{x}^{2}} - \lambda {{x}^{3}} + \alpha {{y}^{2}} + b{{z}^{2}} + c_{b}^{2} \right],
\end{equation}
while
\begin{equation}
{{V}_{h}}\left(x,y,z\right) = \frac{-{{M}_{h}}}{{{\left( {{x}^{2}} + {{y}^{2}} + {{z}^{2}} + c_{h}^{2} \right)}^{1/2}}}.
\end{equation}

Potential (2) describes a triaxial elliptical galaxy with a bulge and a small asymmetry introduced by the term $-\lambda x^3, \lambda << 1$ (see Binney \& Tremaine, 2008). The parameters $\alpha$ and $b$ describe the geometry of the galaxy, while $c_b$ is the scale length of the bulge of the galaxy. The parameter $\upsilon _0$ is used for the consistency of the galactic units. To this potential, we add a spherically symmetric dark halo, described by the potential (3). Here $M_h$ and $c_h$ are the mass and the scale length of the dark halo component respectively.

The aim of this article is: (i) To investigate the character of orbits in the potential (1) and to determine the role played by the dark halo. (ii) To introduce, use and check a new and fast detector, the total angular momentum $L_{tot}$, in order to obtain a reliable criterion to distinguish between order and chaos in galactic potentials.

The Hamiltonian to the potential (1) writes
\begin{equation}
H = \frac{1}{2}\left(p_x^2 + p_y^2 + p_z^2\right) + V_t(x,y,z) = h_3,
\end{equation}
where $p_x$, $p_y$ and $p_z$ are the momenta per unit mass conjugate to $x$, $y$ and $z$ respectively, while $h_3$ is the numerical value of the Hamiltonian.

In this article, we use a system of galactic units, where the unit of length is 1 kpc, the unit of mass is 2.325 $\times$ $10^7$ M$_{\odot}$   and the unit of time is 0.97748 $\times$ $10^8$ yr. The velocity unit is 10 km/s, while $G$ is equal to unity. The energy unit (per unit mass) is 100 (km/s)$^2$. In the above units we use the values: $\upsilon _0 = 15, c_b = 2.5, \alpha = 1.5, b = 1.8, \lambda = 0.03$, while $M_h$ and $c_h$ are treated as parameters.
\begin{figure*}[!tH]
\centering
\resizebox{0.90\hsize}{!}{\rotatebox{0}{\includegraphics*{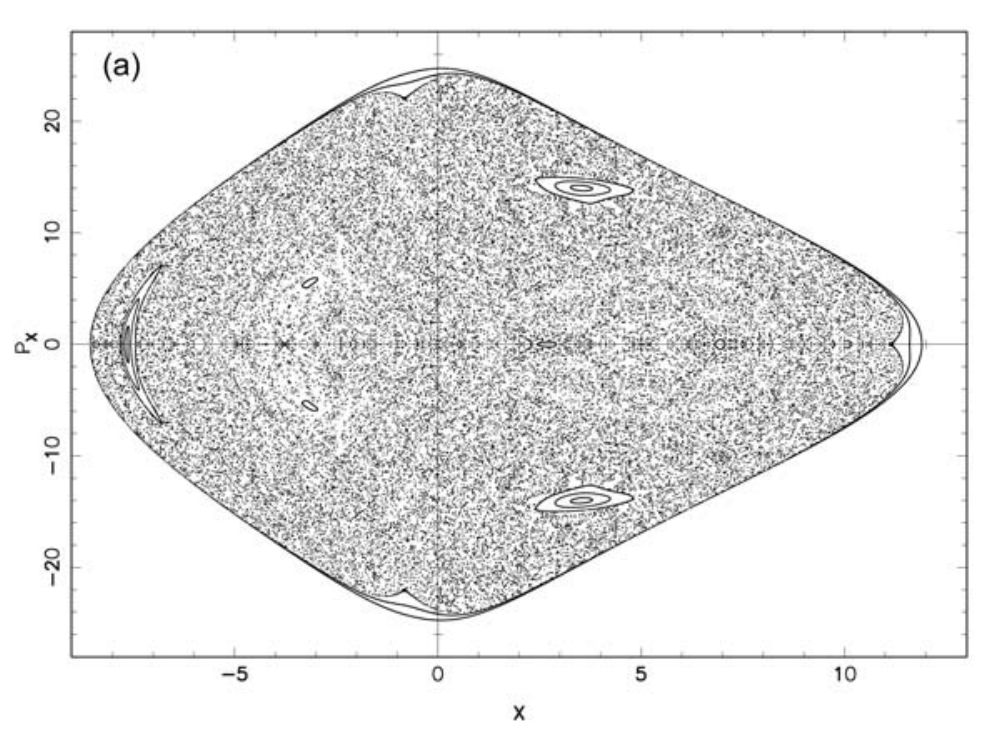}}\hspace{1cm}
                          \rotatebox{0}{\includegraphics*{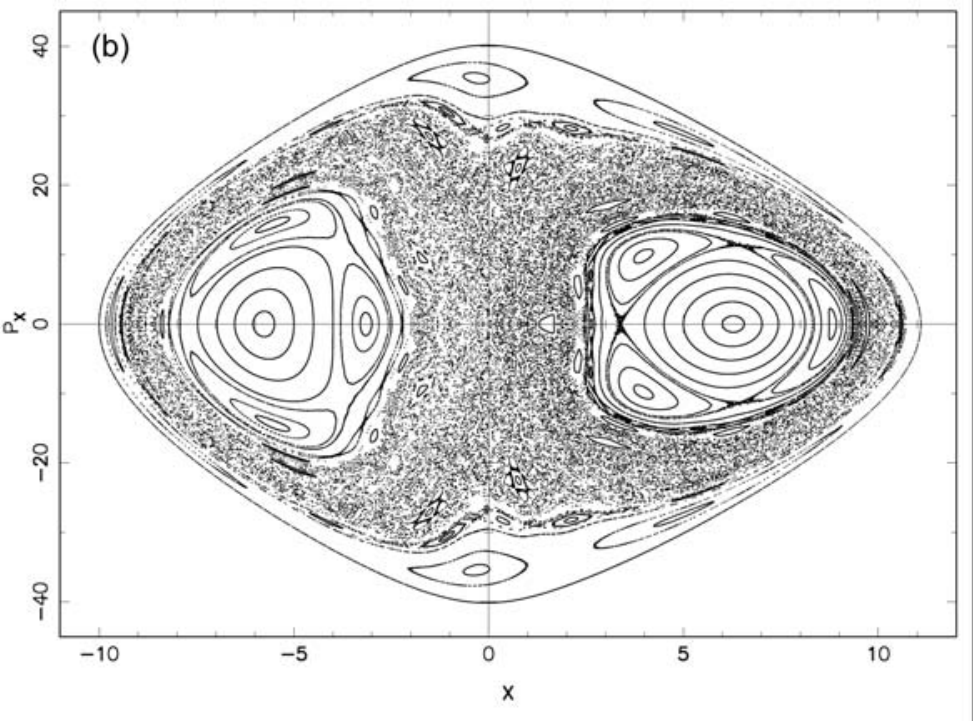}}}
\resizebox{0.90\hsize}{!}{\rotatebox{0}{\includegraphics*{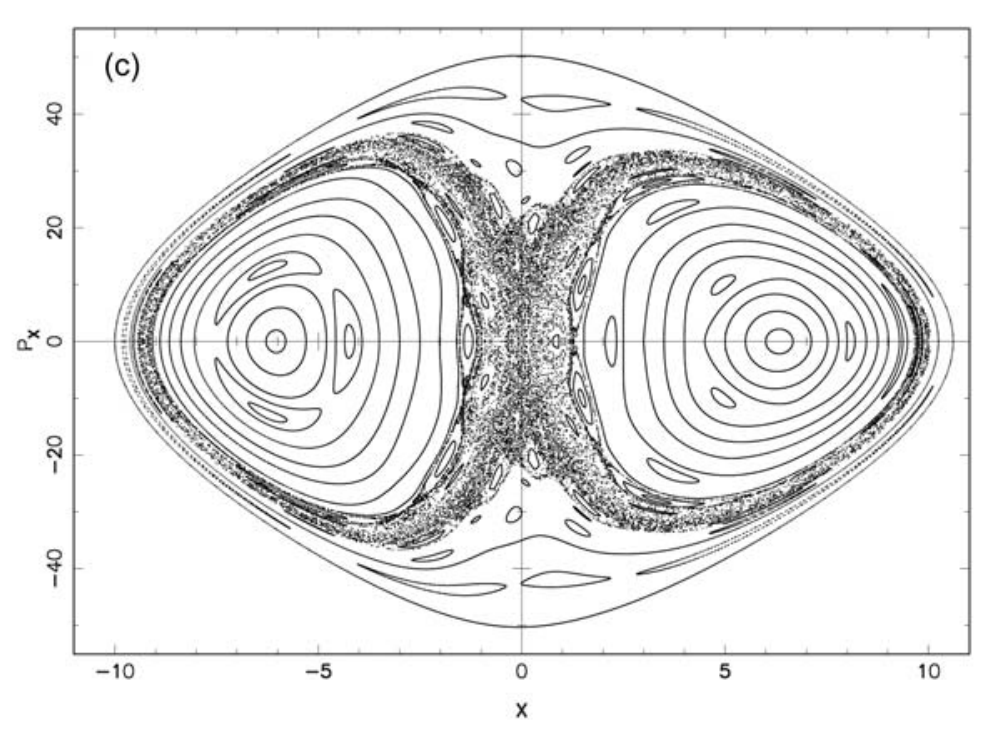}}\hspace{1cm}
                          \rotatebox{0}{\includegraphics*{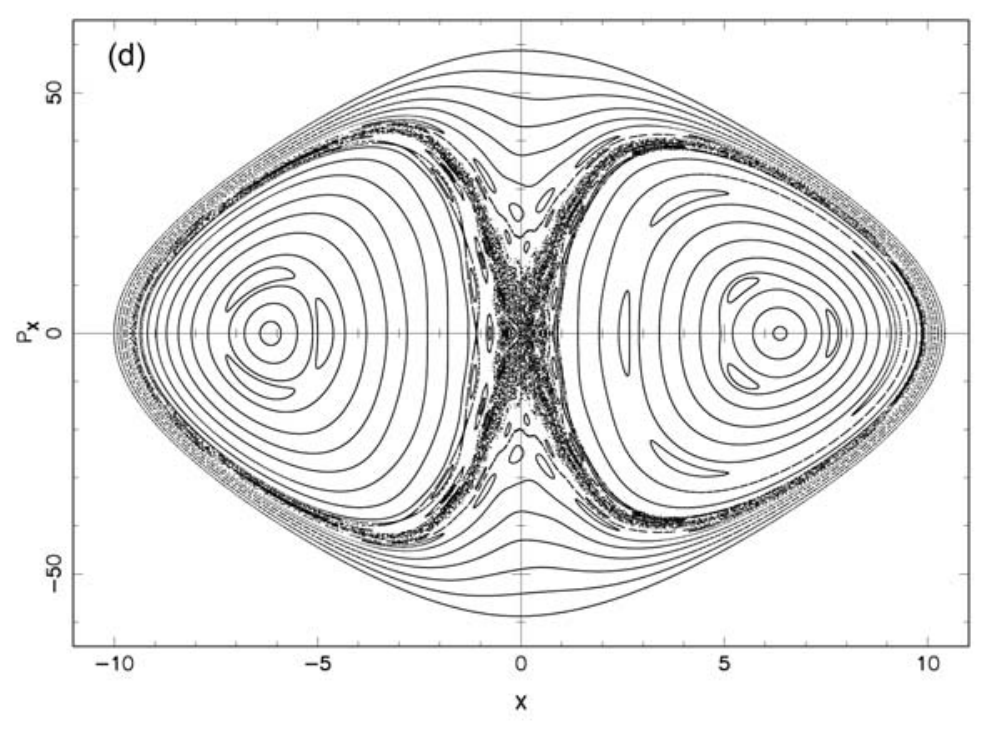}}}
\vskip 0.01cm
\caption{(a-d): The $(x,p_x)$ phase plane, when (a, \textit{upper left}): $M_h = 0, h_2 = 516$, (b, \textit{upper right}): $M_h = 10000, h_2 = -226$, (c, \textit{lower left}): $M_h = 20000, h_2 = -1007$ and (d, \textit{lower right}): $M_h = 30000, h_2 = -1788$. The values of all other parameters are given in text.}
\end{figure*}
\begin{figure*}[!tH]
\centering
\resizebox{0.90\hsize}{!}{\rotatebox{0}{\includegraphics*{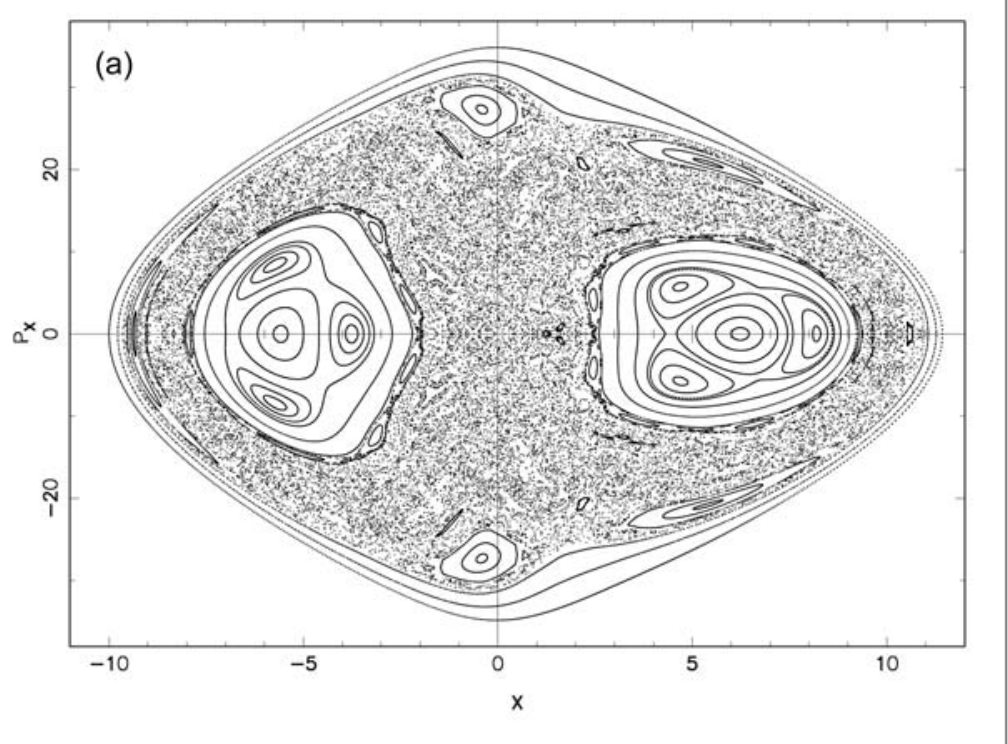}}\hspace{1cm}
                          \rotatebox{0}{\includegraphics*{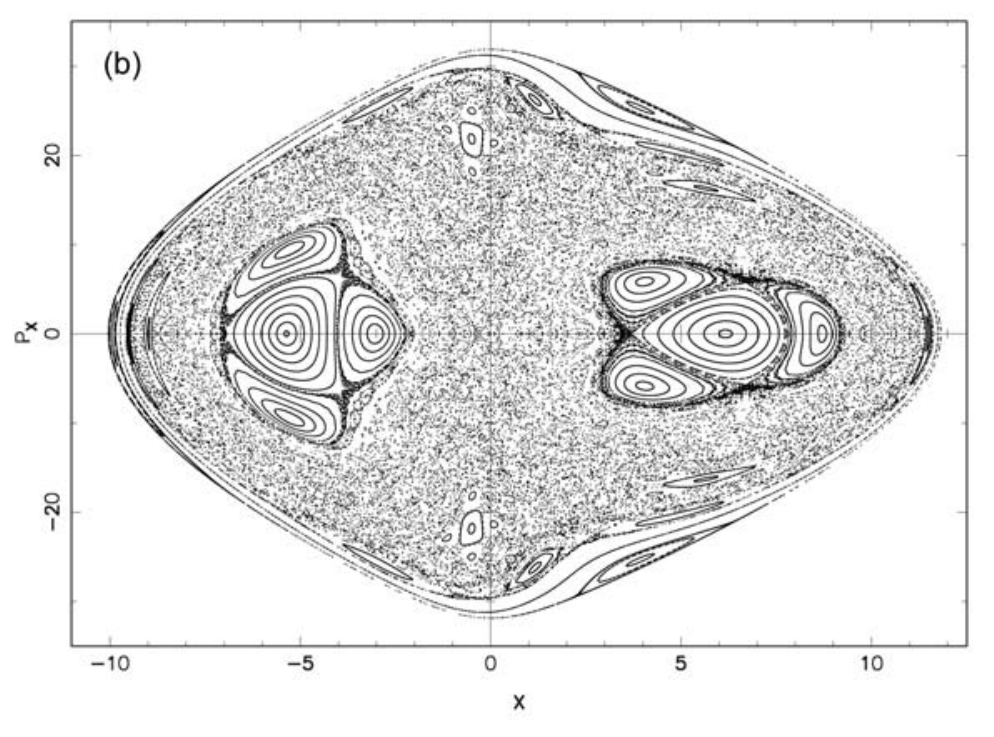}}}
\resizebox{0.90\hsize}{!}{\rotatebox{0}{\includegraphics*{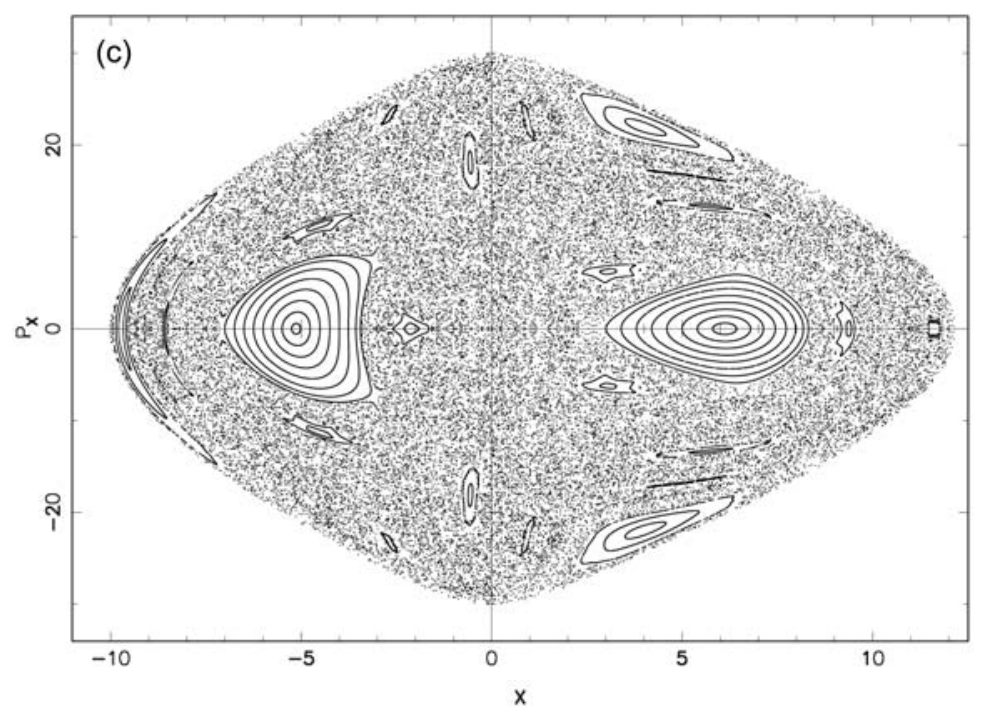}}\hspace{1cm}
                          \rotatebox{0}{\includegraphics*{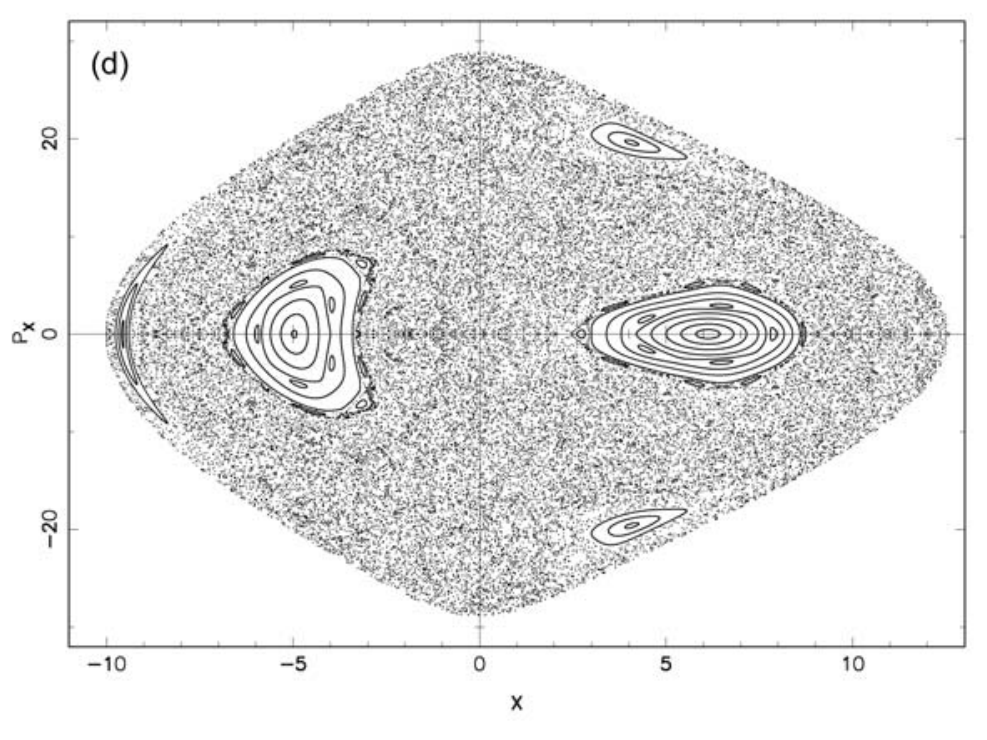}}}
\vskip 0.01cm
\caption{(a-d): The $(x,p_x)$ phase plane, when $M_h = 10000$ and (a, \textit{upper left}): $c_h = 10.5, h_2 = -135$, (b, \textit{upper right}): $c_h = 13, h_2 = -55$, (c, \textit{lower left}): $c_h = 15.5, h_2 = 11$ and (d, \textit{lower right}): $c_h = 18, h_2 = 68$. The values of all other parameters are given in text.}
\end{figure*}
\begin{figure*}[!tH]
\centering
\resizebox{0.90\hsize}{!}{\rotatebox{0}{\includegraphics*{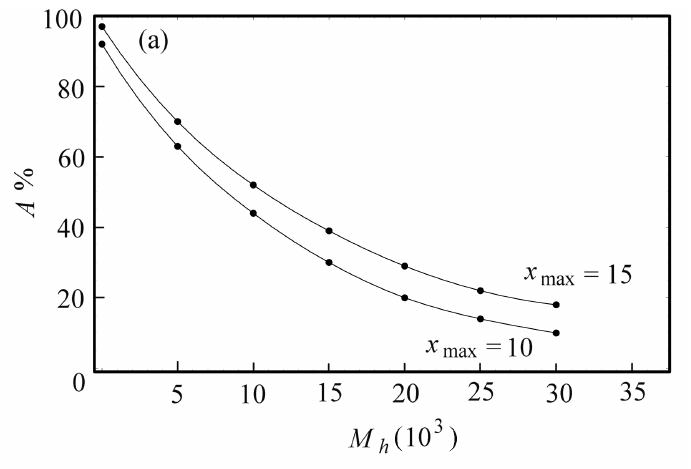}}\hspace{1cm}
                          \rotatebox{0}{\includegraphics*{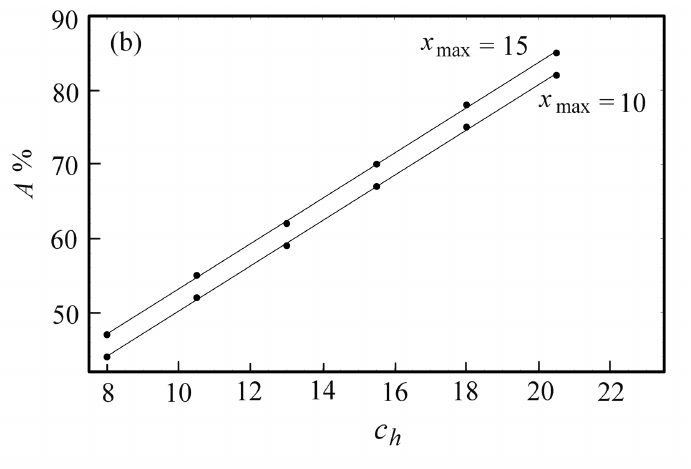}}}
\vskip 0.01cm
\caption{(a-b): (a, \textit{left}): A plot of the area $A\%$ covered by chaotic orbits versus $M_h$ and (b, \textit{right}): A plot of the area $A\%$ covered by chaotic orbits versus $c_h$. The values of all the other parameters are given in text.}
\end{figure*}
\begin{figure*}[!tH]
\centering
\resizebox{0.90\hsize}{!}{\rotatebox{0}{\includegraphics*{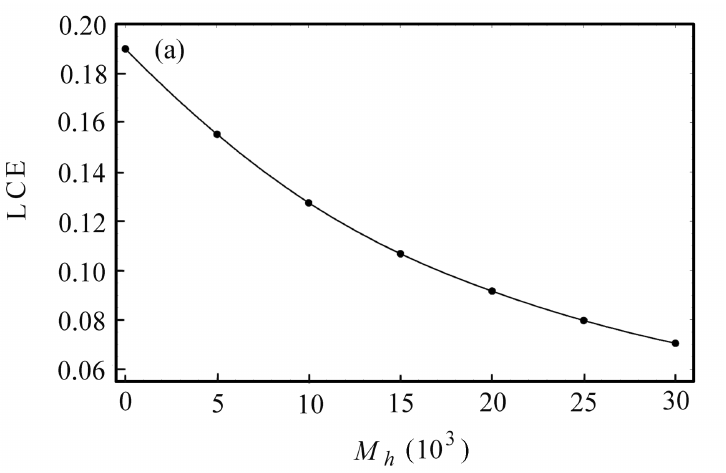}}\hspace{1cm}
                          \rotatebox{0}{\includegraphics*{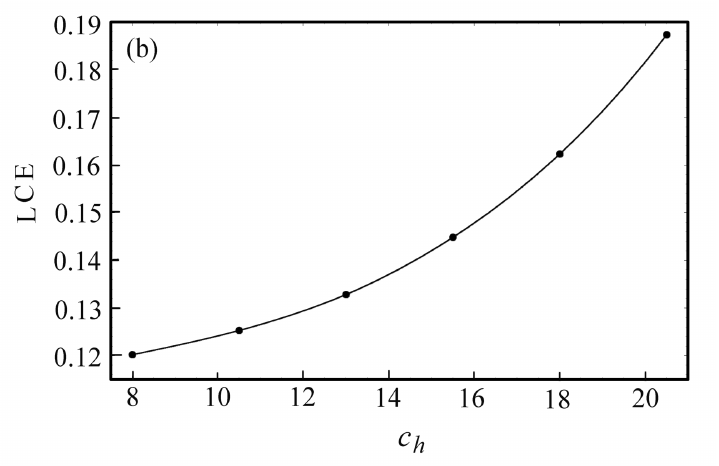}}}
\vskip 0.01cm
\caption{(a-b): (a, \textit{left}): A plot of the L.C.E versus $M_h$ and (b, \textit{right}): A plot of the L.C.E versus $c_h$. The values of all the other parameters are given in text.}
\end{figure*}

\section{Results for the 2D system}

The total angular momentum for a star of unit mass, moving in a 3D orbit is
\begin{equation}
L_{tot} = \sqrt{L_x^2 + L_y^2 + L_z^2},
\end{equation}
where $L_x$ $L_y$ and $L_z$ are the three components of the angular momentum along the $x$, $y$ and $z$ axis given by
\begin{eqnarray}
L_x &=& y\dot{z} - \dot{y}z, \nonumber \\
L_y &=& z\dot{x} - \dot{z}x, \nonumber \\
L_z &=& x\dot{y} - \dot{x}y,
\end{eqnarray}
where the dot indicates derivative with respect to the time. For a 2D system we set in (6), $z=\dot{z}=0$, that is $L_{tot}$ reduces to $L_z$. In this research, we shall use the plot of the $L_{tot}$ vs time in order to distinguish regular from chaotic motion.

Our next step is to study the properties of the 2D dynamical system, which comes from potential, (1) if we set $z = 0$. The corresponding 2D Hamiltonian writes
\begin{equation}
H_2 = \frac{1}{2}\left(p_x^2 + p_y^2\right) + V_t(x,y) = h_2,
\end{equation}
where $h_2$ is the numerical value of the Hamiltonian. We do this in order to use the results obtained for the 2D model in the study of the more complicated 3D model, which will be presented in the next Section.
\begin{figure*}[!tH]
\centering
\resizebox{0.90\hsize}{!}{\rotatebox{0}{\includegraphics*{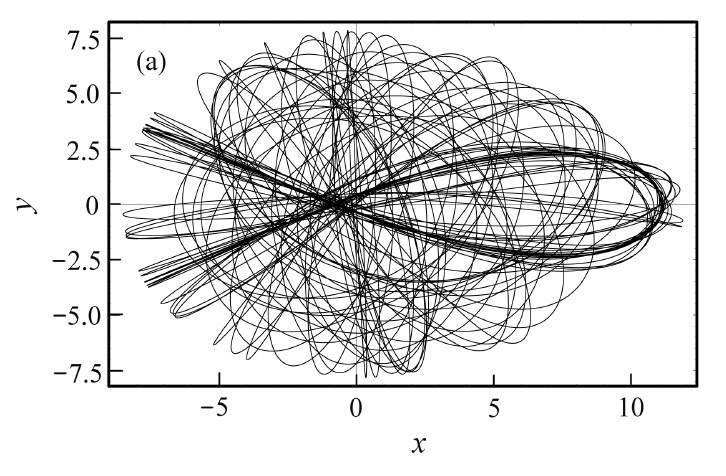}}\hspace{1cm}
                          \rotatebox{0}{\includegraphics*{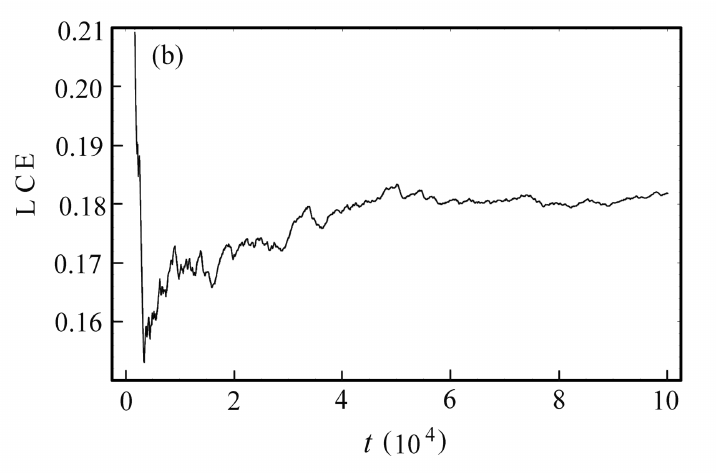}}}
\resizebox{0.90\hsize}{!}{\rotatebox{0}{\includegraphics*{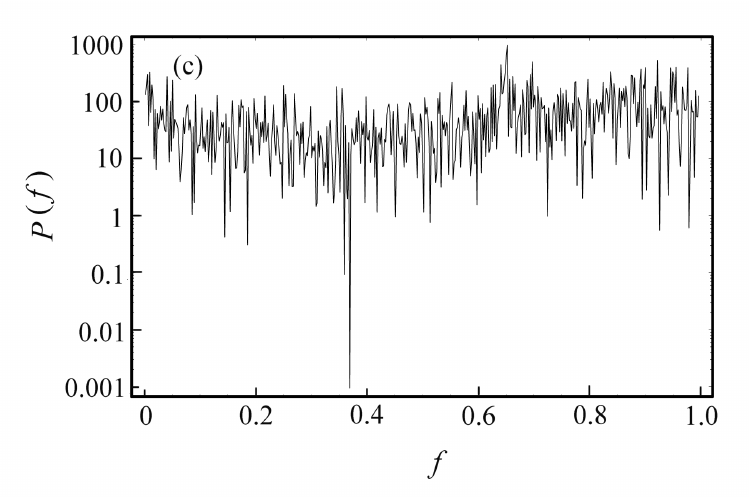}}\hspace{1cm}
                          \rotatebox{0}{\includegraphics*{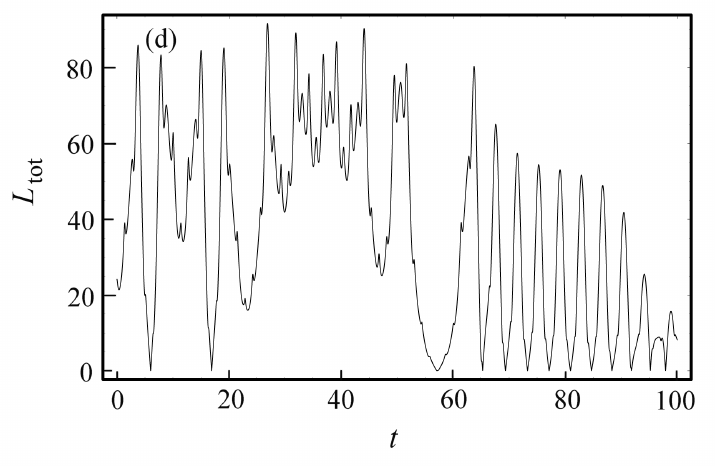}}}
\vskip 0.01cm
\caption{(a-d): (a, \textit{upper left}): An orbit in the 2D potential, (b, \textit{upper right}): The corresponding L.C.E,
(c, \textit{lower left}): The $P(f)$ indicator and (d, \textit{lower right}): The $L_{tot}$ indicator. The motion in chaotic. See text for details.}
\end{figure*}
\begin{figure*}[!tH]
\centering
\resizebox{0.90\hsize}{!}{\rotatebox{0}{\includegraphics*{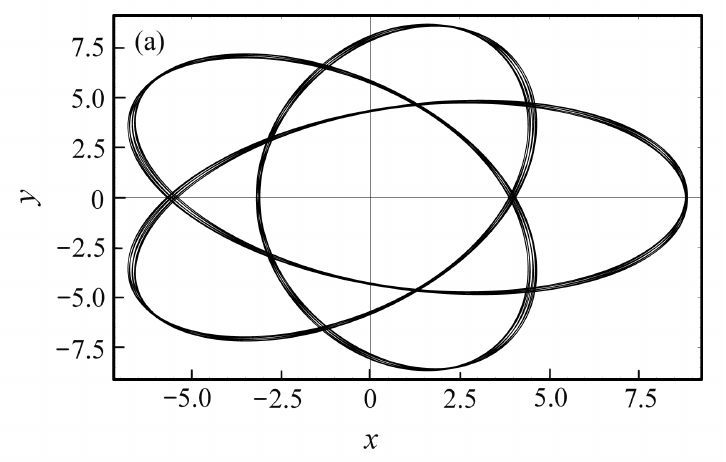}}\hspace{1cm}
                          \rotatebox{0}{\includegraphics*{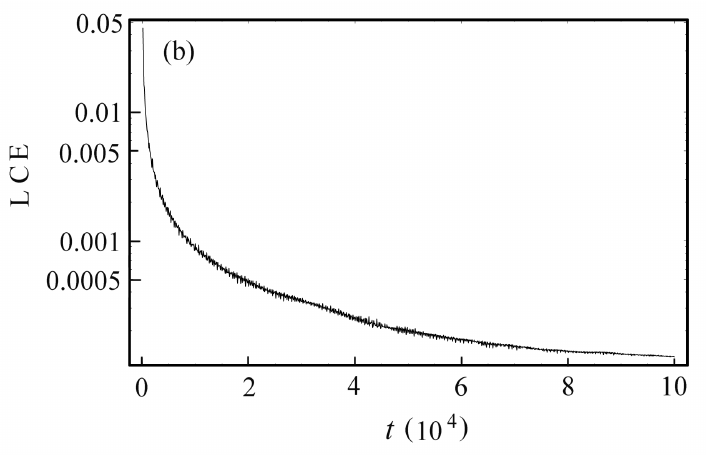}}}
\resizebox{0.90\hsize}{!}{\rotatebox{0}{\includegraphics*{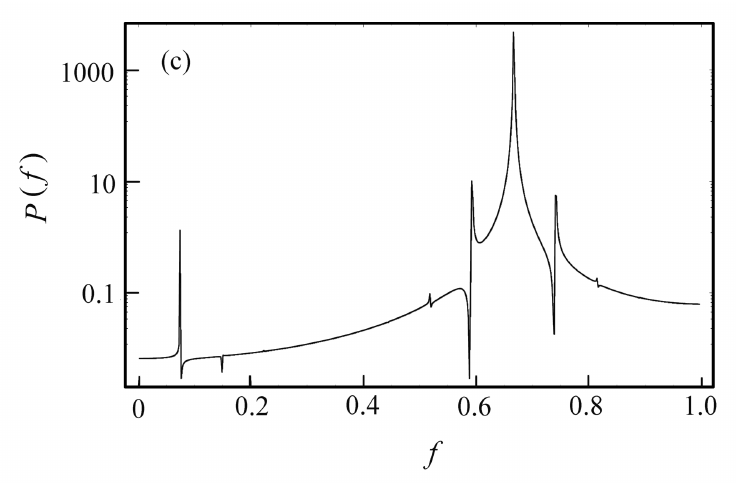}}\hspace{1cm}
                          \rotatebox{0}{\includegraphics*{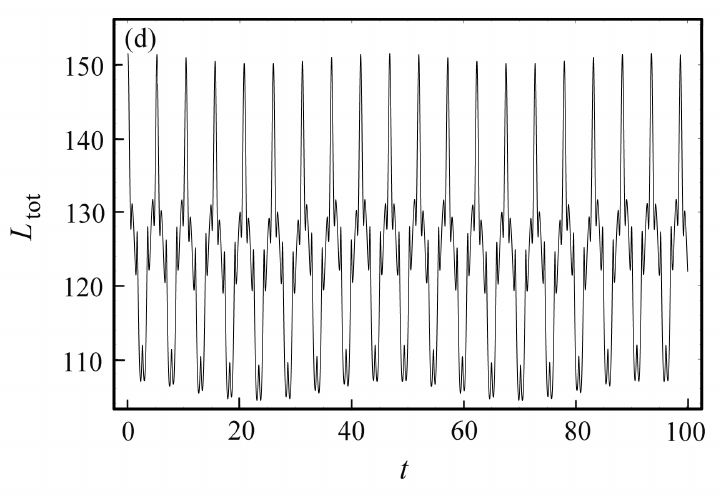}}}
\vskip 0.01cm
\caption{(a-d): Similar as Fig. 5a-d. The motion is regular.}
\end{figure*}
\begin{figure*}[!tH]
\centering
\resizebox{0.90\hsize}{!}{\rotatebox{0}{\includegraphics*{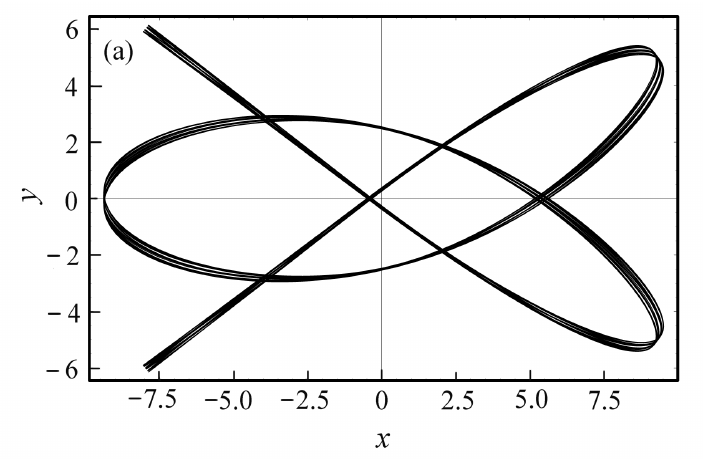}}\hspace{1cm}
                          \rotatebox{0}{\includegraphics*{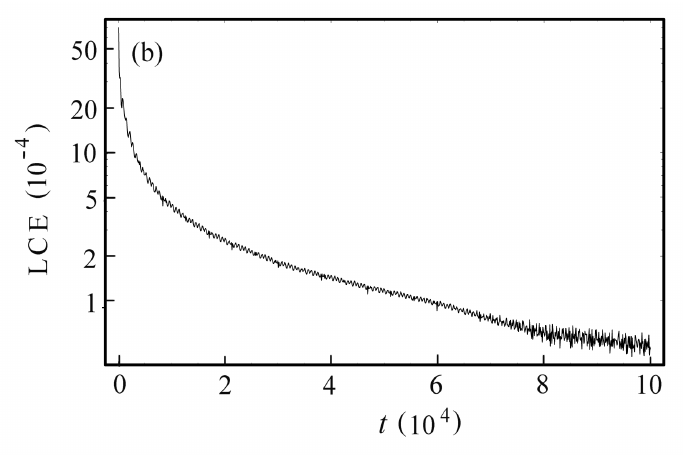}}}
\resizebox{0.90\hsize}{!}{\rotatebox{0}{\includegraphics*{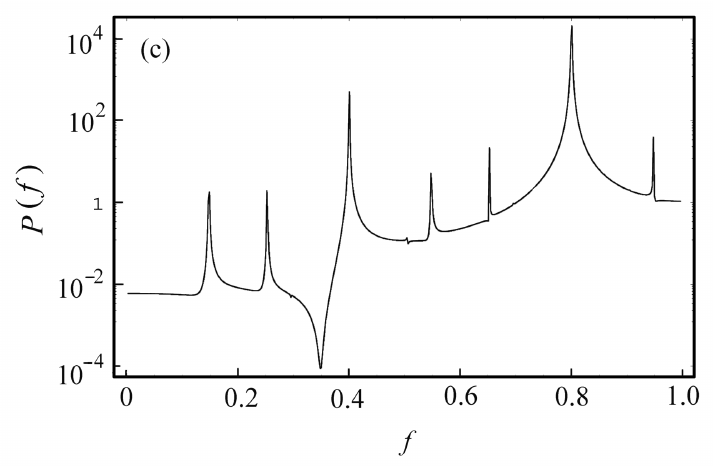}}\hspace{1cm}
                          \rotatebox{0}{\includegraphics*{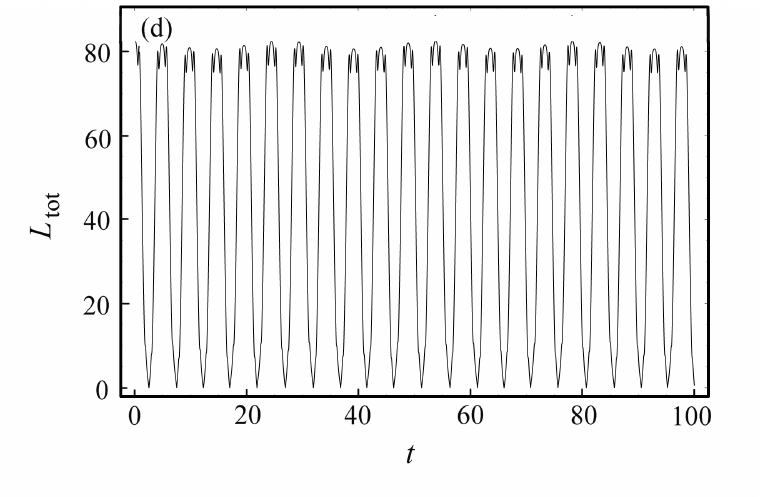}}}
\vskip 0.01cm
\caption{(a-d): Similar as Fig. 6a-d but for a regular orbit.}
\end{figure*}
\begin{figure*}[!tH]
\centering
\resizebox{0.90\hsize}{!}{\rotatebox{0}{\includegraphics*{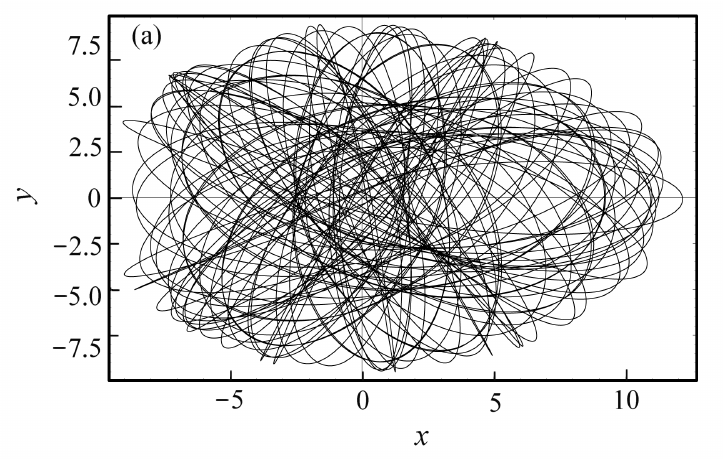}}\hspace{1cm}
                          \rotatebox{0}{\includegraphics*{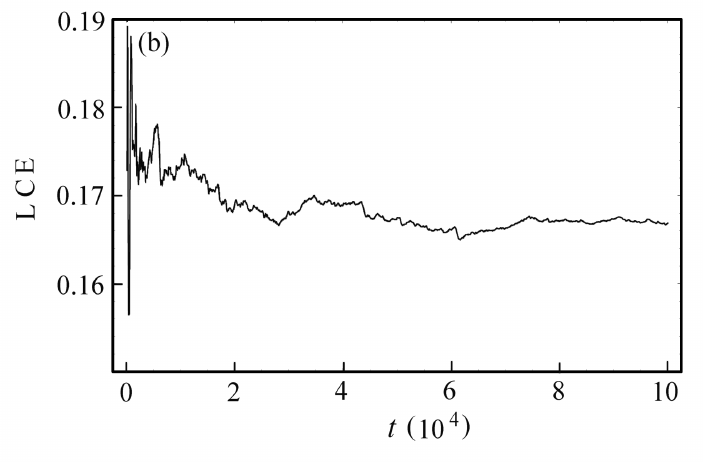}}}
\resizebox{0.90\hsize}{!}{\rotatebox{0}{\includegraphics*{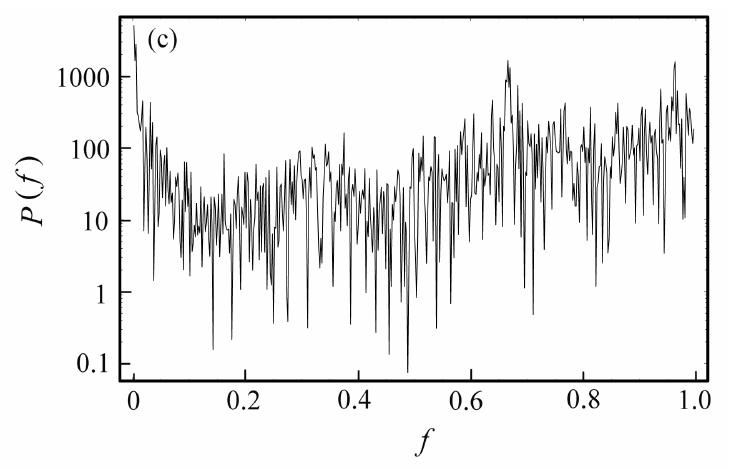}}\hspace{1cm}
                          \rotatebox{0}{\includegraphics*{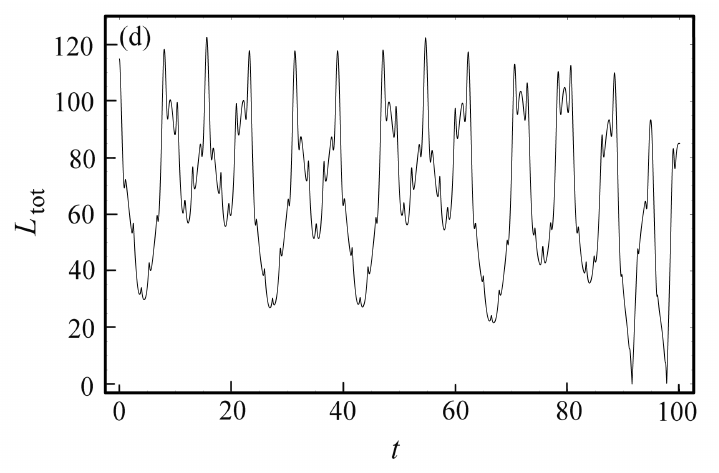}}}
\vskip 0.01cm
\caption{(a-d): Similar as Fig. 5a-d but for a chaotic orbit.}
\end{figure*}

Figure 1a-d shows the $(x,p_x)$, $\left(y = 0, p_y > 0\right)$ phase plane, for four different values of the mass of the dark halo. The values of all the other parameters are: $\upsilon _0 = 15, c_b = 2.5, \alpha = 1.5, b = 1.8, \lambda = 0.03$ and $c_h = 8$. Fig. 1a shows the phase plane, when the system has no halo component, that is when $M_h = 0$. The value of $h_2$ is 516. Fig. 1b is similar to Fig. 1a but when $M_h = 10000$ and $h_2 = - 226$. Fig. 1c is similar to Fig. 1a but when $M_h = 20000$ and $h_2 = -1007$. Fig. 1d shows the phase plane when $M_h = 30000$ and $h_2 = -1788$.
\begin{figure*}[!tH]
\centering
\resizebox{0.90\hsize}{!}{\rotatebox{0}{\includegraphics*{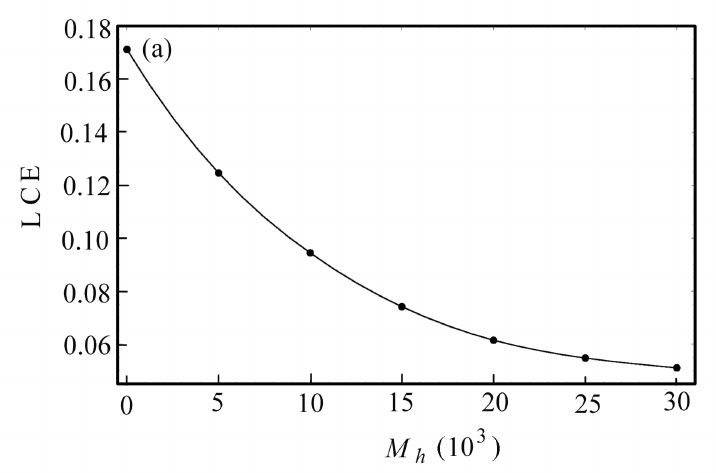}}\hspace{1cm}
                          \rotatebox{0}{\includegraphics*{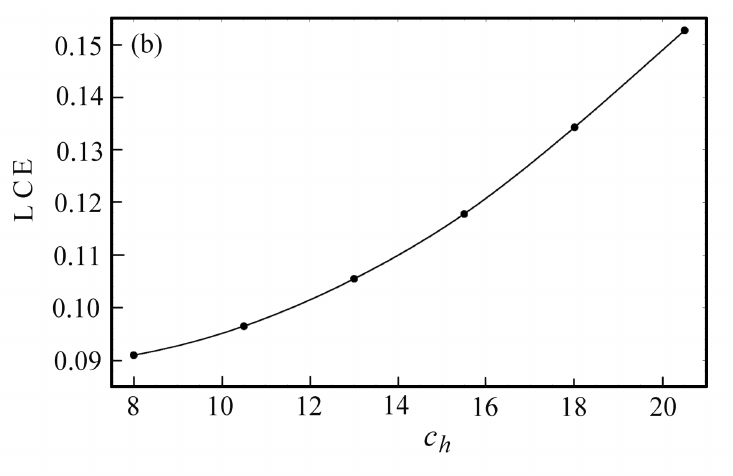}}}
\vskip 0.01cm
\caption{(a-b): Similar to Fig. 4a-b but for the 3D potential. The values of all other parameters are given in text.}
\end{figure*}
\begin{figure*}[!tH]
\centering
\resizebox{0.85\hsize}{!}{\rotatebox{0}{\includegraphics*{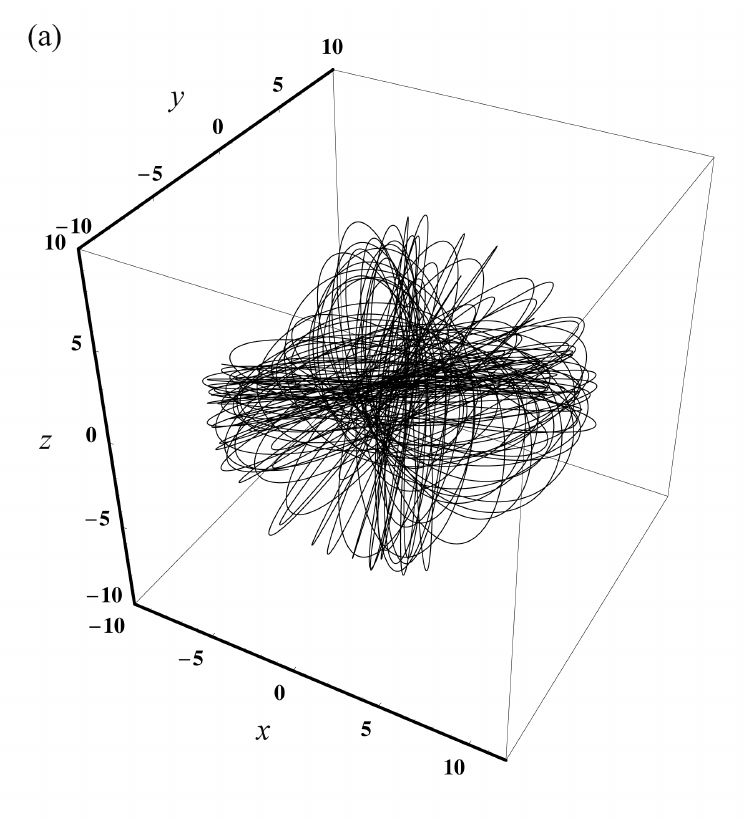}}\hspace{1cm}
                          \rotatebox{0}{\includegraphics*{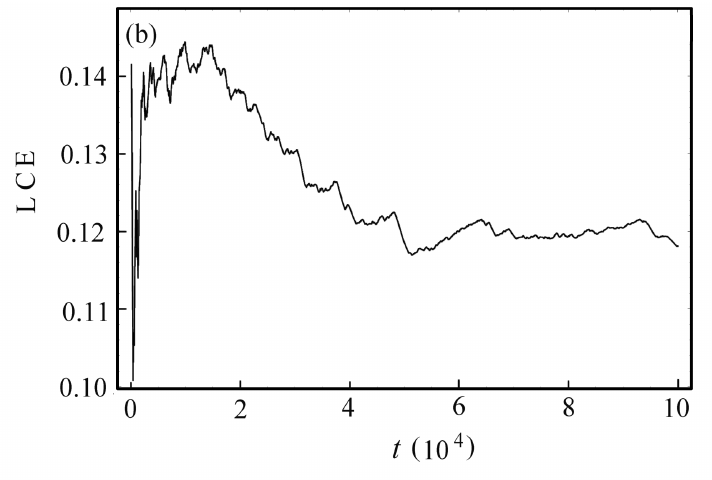}}}
\resizebox{0.85\hsize}{!}{\rotatebox{0}{\includegraphics*{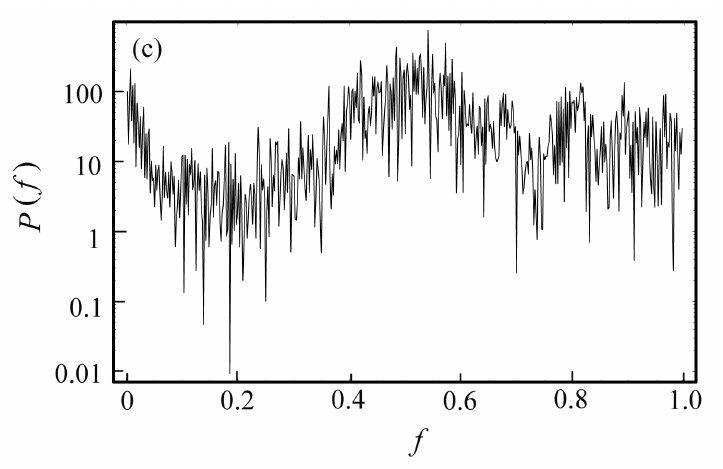}}\hspace{1cm}
                          \rotatebox{0}{\includegraphics*{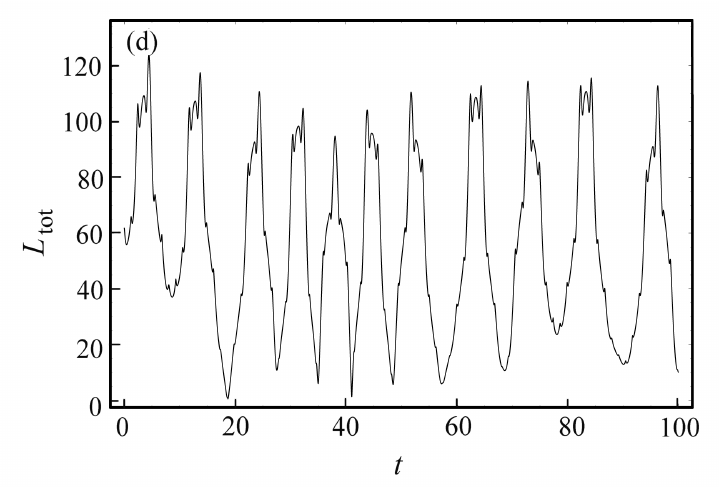}}}
\vskip 0.01cm
\caption{(a-d): (a, \textit{upper left}): An orbit in the 3D potential, (b, \textit{upper right}): The corresponding L.C.E,
(c, \textit{lower left}): The $P(f)$ indicator and (d, \textit{lower right}): The $L_{tot}$ indicator. The motion in chaotic. See text for details.}
\end{figure*}
\begin{figure*}[!tH]
\centering
\resizebox{0.85\hsize}{!}{\rotatebox{0}{\includegraphics*{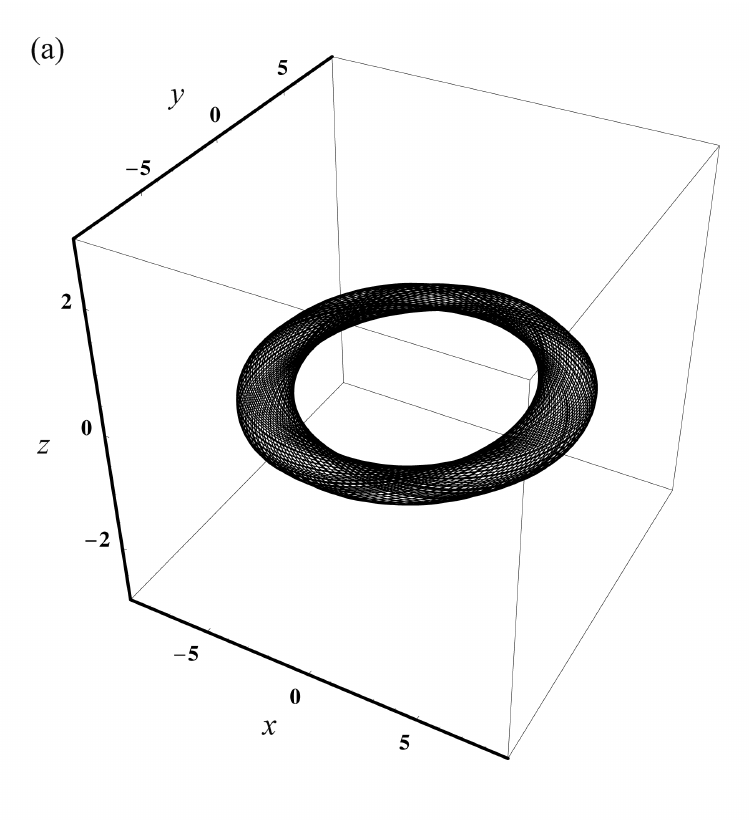}}\hspace{1cm}
                          \rotatebox{0}{\includegraphics*{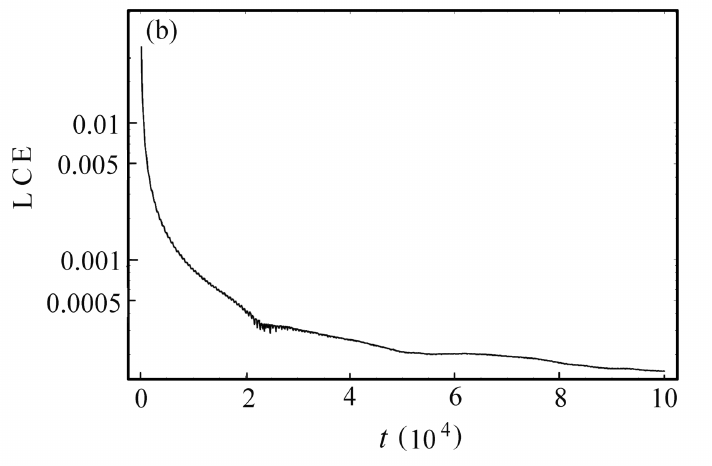}}}
\resizebox{0.85\hsize}{!}{\rotatebox{0}{\includegraphics*{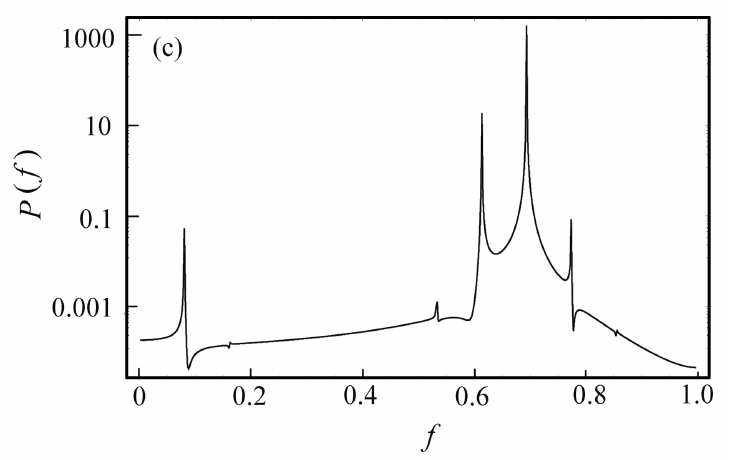}}\hspace{1cm}
                          \rotatebox{0}{\includegraphics*{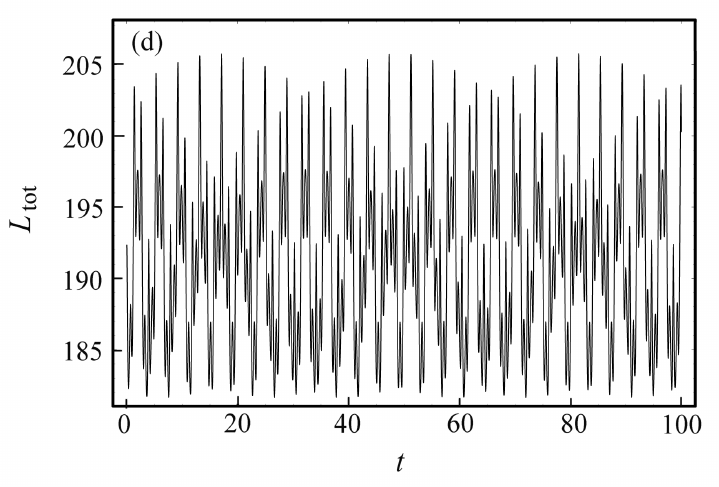}}}
\vskip 0.01cm
\caption{(a-d): Similar as Fig. 10a-d. The motion is regular.}
\end{figure*}

Fig. 2a-d is similar to Fig. 1a-d but when $M_h$ is 10000, while $c_h$ is treated as a parameter. All the other parameters are as in Fig. 1. In Fig. 2a we have $c_h = 10.5$ and $h_2 = -135$. In Figure 2b the values of $c_h$ and $h_2$ are 13 and -55 respectively. In the phase plane shown in Fig. 2c, we have taken $c_h = 15.5$ and $h_2 = 11$. In Fig. 2d we have chosen $c_h = 18$ and $h_2 = 68$.

Figure 3a shows the percentage of the phase plane $A\%$ covered by chaotic orbits as a function of the mass of the dark halo, for two different values of $x_{max}$. The values of the parameters are: $\upsilon_0 = 15, c_b = 2.5, \alpha = 1.5, b = 1.8, \lambda = 0.03$ and $c_h = 8$. Figure 3b shows a plot between $A\%$ and $c_h$. The values of the parameters are: $\upsilon_0 = 15, c_b = 2.5, \alpha = 1.5, b = 1.8, \lambda = 0.03, M_h = 10000$. Figures 4a and 4b show a plot of the Lyapunov Characteristic Exponent (L.C.E) versus $M_h$ or versus $c_h$ respectively.

In what follows, we shall investigate the regular or chaotic character of orbits in the 2D Hamiltonian (7) using the new dynamical detector $L_{tot}$ . In order to see the effectiveness of the new method, we shall compare the results with two other indicators, the classical method of the L.C.E and the $P(f)$ spectral method, used by Karanis \& Vozikis (2007). This method uses the Fast Fourier Transform (F.F.T) of a series of time intervals, each one representing the time that elapsed between two successive points on the Poincar\'{e} $(x,p_x)$ phase plane for 2D systems, while for 3D systems they take two successive points on the plane $z = 0$.

Figure 5a shows an orbit with initial conditions: $x_0 = -1.0, y_0 = p_{x0} = 0$, while the value of $p_{y0}$ is always found from the energy integral for all orbits. The values of all the other parameters and the energy are as in Fig. 1a. One observes in Figure 5b that the L.C.E, which was computed for a period of $10^5$ time units, has a value of about 0.18 indicating chaotic motion. The same result is shown by the $P(f)$ indicator, which is given in Figure 5c. Figure 5d shows a plot of the $L_{tot}$ vs time, for a time interval of 100 time units. We see that the diagram is highly asymmetric, with a number of small and large peaks. The above characteristics suggest that the corresponding orbit is chaotic.

Figure 6a shows an orbit with initial conditions: $x_0 = 8.8, y_0 = p_{x0} = 0$. The values of all the other parameters and the energy are as in Fig. 1b. As we see, this is a quasi periodic orbit. Therefore the L.C.E of this orbit goes to zero, as it is clearly seen in Figure 6b. The $P(f)$ indicator in Figure 6c shows a small number of peaks, also indicating regular motion. The plot of the $L_{tot}$ given in Figure 6d is now quasi periodic, with symmetric peaks, indicating regular motion.

Figure 7a-d is similar to Fig. 6a-d for an orbit with initial conditions: $x_0 = -9.36, y_0 = p_{x0} = 0$, while the values of all the other parameters and the energy are as in Fig. 2a. As we see, the orbit is quasi periodic and this fact is indicated by all three dynamical parameters. On the contrary, the orbit shown in Figure 8a has initial conditions: $x_0 = 10, y_0 = p_{x0} = 0$ and the values of all the other parameters and the energy are as in Fig. 2d. The orbit looks chaotic and this is indicated by the L.C.E, the $P(f)$ and the $L_{tot}$ shown in Figs. 8b, 8c and 8d respectively.

A large number of orbits in the 2D system were calculated for different values of the parameters. All numerical results suggested that the $L_{tot}$ is a fast and reliable dynamical parameter and can be safely used in order to distinguish ordered from chaotic motion.

\section{Results for the 3D system}

The regular or chaotic nature of the 3D orbits is found as follows: we choose initial conditions $\left(x_0, p_{x0}, z_0\right)$, $y_0 = p_{z0} = 0$, such as $\left(x_0, p_{x0}\right)$ is a point on the phase plane of the 2D system. The point $\left(x_0, p_{x0}\right)$ lies inside the limiting curve
\begin{equation}
\frac{1}{2}p_x^2 + V_t(x) = h_2,
\end{equation}
which is the curve containing all the invariant curves of the 2D system. We choose $h_3 = h_2$ and the value of $p_{y0}$ for all orbits is obtained from the energy integral (4). Our numerical experiments show that orbits with initial conditions $\left(x_0, p_{x0}, z_0\right)$, $y_0 = p_{z0} = 0$, such as $\left(x_0,p_{x0}\right)$ is a point in the chaotic regions of Figs. 1a-d and 2a-d for all permissible values of $z_0$, produce chaotic orbits.

Our next step is to study the character of orbits with initial conditions $\left(x_0, p_{x0}, z_0\right)$, $y_0 = p_{z0} = 0$, such as $\left(x_0, p_{x0}\right)$ is a point in the regular regions of Figs. 1a-d and  2a-d. It was found, that in all cases the regular or chaotic character of the above 3D orbits depends strongly on the initial value $z_0$. Orbits with small of $z_0$ are regular, while for large values of $z_0$ they change their character and become chaotic. The general conclusion, which is based on the results derived from a large number of orbits, is that orbits with values of $z_0 \geq 0.75$ are chaotic, while orbits with values of $z_0 < 0.75$ are regular.

Figure 9a shows the L.C.E of the 3D system, as a function of the mass of halo, for a large number of chaotic orbits, when $c_h = 8$. Figure 9b shows he L.C.E as a function of $c_h$ when $M_h = 10000$.

Figure 10a-d is similar to Fig. 8a-d but for a 3D orbit. The orbit shown in Fig. 10a looks chaotic. The initial conditions are: $x_0 = 2.0, p_{x0} = 0, z_0 = 0.5$. Remember that all orbits have $y_0 = p_{z0} = 0$, while the value of $p_{y0}$ is always found from the energy integral. The values of all the other parameters and the energy $h_3$ are as in Fig. 2b. The L.C.E shown in Figure 10b assures the chaotic character of the orbit. The $P(f)$ given in Figure 10c also suggests chaotic motion. The same conclusion comes from the $L_{tot}$, which is shown in Figure 10d. Figure 11a-d is similar to Fig. 10a-d but for a quasi periodic 3D orbit. The initial conditions are: $x_0 = 5.0, p_{x0} = 0, z_0 = 0.1$. The values of all the other parameters and the energy $h_3$ are as in Fig. 1c. Here one observes, that all the three detectors support the regular character of orbit.

The conclusion for the study of the 3D model, is that the $L_{tot}$ detector can give reliable and very fast results for the character of the orbits. There is no doubt, that the $L_{tot}$ is faster than the two other indicators, used in this research. Therefore, one can say that this indicator is a very useful tool for a quick study of the character of orbits in galactic potentials.

\section{Discussion and conclusions}

The main conclusions of this research are the following:

\textbf{1.} The percentage of the chaotic orbits decreases as the mass of the spherical halo increases. Therefore, the mass of the dark halo can be considered as an important physical quantity, acting as a controller of chaos in galaxies showing small asymmetries.

\textbf{2.} One expects to observe a smaller fraction of chaotic orbits in asymmetric triaxial galaxies with a dense spherical halo, while the fraction of chaotic orbits would increases in asymmetric triaxial galaxies surrounded by less dense spherical dark halo components.

\textbf{3.} It was found that the L.C.E in both the 2D and the 3D models decreases, as the mass of halo increases, while the L.C.E increases as the scale length $c_h$ of the halo increases. This means that not only the percentage of chaotic orbits, but also the degree of chaos is affected by the mass or the scale length of the spherically symmetric dark halo component.

\textbf{4.} The $L_{tot}$ gives fast and reliable results regarding the nature of motion, both in 2D and 3D galactic potentials. For all calculated orbits the results given by the $L_{tot}$ coincide with the outcomes obtained using the L.C.E or the $P(f)$ spectral method. The advantage of the $L_{tot}$ is that is faster than the above two methods.

\end{document}